\documentclass[12pt,preprint]{aastex}
\usepackage{graphics}
\usepackage{subfigure}
\usepackage{amsmath}
\usepackage{bm}

\shorttitle{Particle acceleration in superluminal strong waves}
\shortauthors{Teraki et al.}

\begin{document}

\title{Particle acceleration in superluminal strong waves}

\author{Yuto Teraki\altaffilmark{1}, Hirotaka Ito\altaffilmark{1} and Shigehiro Nagataki\altaffilmark{1}}
\affil{Astrophysical Big Bang Laboratory, RIKEN, Saitama 351-0198, Japan}
\email{yuto.teraki@riken.jp}

\begin{abstract}
We calculate the electron acceleration in random superluminal strong waves (SLSWs) and radiation from them
by using numerical methods in the context of the termination shock of the pulsar wind nebulae.
We pursue the electrons by solving the equation of motion in the analytically expressed electromagnetic turbulences.
These consist of primary SLSW and isotropically distributed secondary electromagnetic waves.
Under the dominance of the secondary waves, all electrons gain nearly equal energy.
On the other hand, when the primary wave is dominant, selective acceleration occurs.
The phase of the primary wave felt by the electrons moving nearly along the wavevector 
changes very slowly compared to the oscillation of the wave, which is called "phase locked",
and such electrons are continuously accelerated.
This acceleration by SLSWs may play a crucial role in the pre-acceleration for the shock acceleration.
In general, the radiation from the phase-locked population is different from the synchro-Compton radiation.
However, when the amplitude of the secondary waves is not extremely weaker than that of the primary wave,
the typical frequency can be estimated from the synchro-Compton theory by using the secondary waves.
The primary wave does not contribute to the radiation, because the SLSW accelerates electrons almost linearly.
This radiation can be observed as a radio knot at the upstream of the termination shock
of the pulsar wind nebulae without counter parts in higher frequency range.
\end{abstract}


\section{Introduction}
\label{intro}
Particle acceleration is one of the most important physical processes in the astrophysics and plasma physics.
The acceleration by the strong electromagnetic waves (EM waves) has been investigated for a long period in the field of laser physics 
(e.g. Jory and Trivelpiece 1968, Sarachik and Schappert 1970, Karimdabadi et al. 1990, Kuznetsov 2014).
In the field of astrophysics, such mechanisms were investigated mainly in the context of the pulsar. 
The magnetic dipole radiation was thought to be emitted from the pulsar which has inclined magnetic axis 
to the rotation axis (Pacini 1968).
This EM wave was regarded as a "strong wave" in the vicinity of the pulsar.
The strength of the EM wave is generally defined by the strength parameter $a \equiv eE/mc\omega$, where $E$ and $\omega$ are
the amplitude and frequency of the EM wave, $m$ and $e$ are the mass and charge of the electron,
and $c$ is the speed of light in vacuum.
It is estimated to be much larger than unity for the EM wave around the pulsar, which means that 
this EM wave is capable to accelerate the electrons to a relativistic energy.
Electrons dropped in the strong EM wave at rest are strongly accelerated toward the direction of the wavevector.
The phase of the wave felt by the electrons changes very slowly (phase locking), since electron speed becomes close to speed of light.
The strong wave is nearly a stationary EM field for these electrons, and they are continuously accelerated.
If we assume the infinite plane wave, the maximum Lorentz factor is $\gamma_{\rm max}\sim a^2$, not $a$ (Gunn \& Ostriker 1971).
To achieve $\gamma_{\rm max}\sim a^2$, phase locking effect plays a crucial role.

As is well known, the pulsar magnetosphere is not a vacuum as was assumed by the above papers, but is filled with dense plasma 
(Goldreich \& Julian 1969). 
Kegel (1971) pointed out that when the refraction index $n_{\rm r}$ is significantly smaller than unity
because of the existence of the plasma, the phase locking becomes inefficient since the phase velocity $c/n_{\rm r}$ 
of the EM wave becomes significantly larger than $c$.
Such waves are called superluminal strong waves (SLSWs).
Moreover, if the plasma is sufficiently dense, even the propagation near the light cylinder is prohibited.
When we consider the pair creation in the magnetosphere, the plasma around the light cylinder 
is much larger than the Goldreich-Julian density.
The low frequency EM waves such as the dipole radiation from the pulsar cannot propagate in such a high density plasma.
The magnetic energy is not carried by the EM wave but by the entropy wave in the magnetised plasma wind 
(i.e. striped wind in Coroniti 1990).

Recently, the strong EM waves in the pulsar environment again attract many interests.
It is pointed out that the entropy mode can be converted to the strong EM wave in the outer region of the pulsar wind 
(Arka \& Kirk 2012).
Moreover, it is numerically shown by using relativistic two-fluid simulation that this conversion can occur by the interaction 
with the termination shock (Amano \& Kirk 2013).
Such EM waves have superluminal phase velocity and their strength parameter $a$ may be larger than $1$.
Around the termination shock of the pulsar wind, the electrons are accelerated by shock crossing and radiate synchrotron photons.
If such SLSWs exist around the shock, they may affect the particle acceleration and radiation.

There are many unsolved problems for the particle acceleration around the termination shock of the pulsar wind nebulae.
One is the injection problem.
In general, to be injected into the shock crossing cycle,
the particles have to be supra-thermal when they encounter the shock front. 
However, in the paper of Kennel \& Coroniti (1984a), upstream plasma is assumed to be cold and all particles are accelerated.
In other words, they assumed a very high injection rate.
The lowest Lorentz factor $\gamma \sim 10^6$ at the immediate downstream of the shock corresponds to 
the bulk Lorentz factor of the upstream wind $\Gamma \sim 10^6$.
These particles emit optical photons by the synchrotron radiation.
Their model can explain the observed spectrum at frequency region higher than the optical range, but
radio components were not discussed (Kennel Coroniti 1984b).
There are some models for Crab Nebula which explain the radio components.
For example, we have time-dependent one zone models (Tanaka \& Takahara 2010, Bucciantini, Arons, \& Amato 2011)
and axisymmetric two-dimensional MHD (MagnetoHydroDynamics) models (Olmi et al. 2014, 2015).
They reproduces the spectrum of the Crab Nebula by assuming the energy distribution of the electrons as a broken power law,
but they do not specify the origin of the break Lorentz factor $\gamma \sim 10^6$.
It may imply that there exists a pre-acceleration mechanism in the upstream of the termination shock.
That is to say, we can consider a model in which the pre-acceleration mechanism makes the energy distribution broader, and 
only higher energy components are injected to the shock crossing cycle.
If the SLSWs can make a non-thermal energy distribution in the upstream,
it can be a good candidate of the pre-accelerator.

Radiation from the electron in a strong ($a>1$) EM wave is usually called synchro-Compton radiation (Rees 1971) or 
nonlinear inverse Compton (NIC) radiation (Gunn \& Ostriker 1971).
The deflection angle in one cycle of the electron motion in the interaction with a strong wave 
tends to be larger than $1/\gamma$.
As a result, the radiation signature resembles the synchrotron radiation.
However, we do not know the radiation spectra when there are many waves.
Moreover, even for the interaction with one SLSW, there are ambiguous points in the consideration of the typical frequency.
To estimate the typical frequency of the radiation, we have to know the photon formation time
(Akhiezer \& Shul'ga 1987, Reville \& Kirk 2010), which is the inverse of the cyclotron frequency $mc/eB$ 
in the context of the synchrotron radiation.
Previous studies considered the cases for which the phase locking effect is weak, which means that 
the velocity is oblique to the wavevector direction.
For this case, the photon formation time is $\sim mc/eE$.
In general, the motion of the electron in the strong wave can result in a different photon formation time.
Therefore, the resultant radiation spectra can be different from the NIC or synchro-Compton theories.

In this paper, we study the electron acceleration in the SLSWs.
Also, the radiation signature in such situation is studied.
We use numerical methods to investigate such highly nonlinear motions.
The contents of this paper are as follows.
In section \ref{form}, we discuss the physical parameters for the computational study and 
describe the methods we use. The results are shown in section \ref{results}. 
Section \ref{discussion} presents discussions including observational features. 
We finally summarize this paper in section \ref{summary}.

\section{Formulation}
\label{form}
\subsection{Parameters}
Here, we estimate the physical parameters around the termination shock of the pulsar wind nebula
by using the Crab Pulsar \& its nebula as a representative system.
First, we estimate the strength parameter $a$ of the entropy mode in the immediate upstream of the termination shock
by following the estimation by Kirk \& Mochol (2011), in which they estimated $a$ in the (Active Galactic nuclei) AGN jets.
We note that this strength parameter is not identical to the strength parameter of the EM wave.
The strength parameter of the entropy mode is defined by the wavelength $\lambda_{\rm sw}$ and magnetic field strength $B$
as $eB\lambda_{\rm sw}/2\pi mc^2$ in the observer frame.
There are two feasible assumptions.
One is that the magnetic field strength in pulsar wind is inversely proportional to the distance 
from the pulsar $r$ as $B\propto r^{-1}$,
(here we implicitly assumed the magnetic field configuration to be pure toroidal).
The other is that the spindown luminosity of the pulsar $L_{\rm sd}$ is carried by the Poynting flux (high sigma) and to be isotropic.
On the above assumptions, the entropy wave in the observer frame resembles EM wave due to the fact 
that electric field is perpendicular to the magnetic field, the strength ratio is nearly unity, and the wavevector
is perpendicular to the electric field and magnetic field.
We obtain $a$ from the observed quantities as
\begin{equation}
a= \left(\frac{r_{\rm LC}}{r}\right)\left(\frac{e^2L_{\rm sd}}{m^2c^5}\right)^{1/2}\simeq 3.4\times10^{10} 
  \left(\frac{r_{\rm LC}}{r}\right)\left(\frac{L_{\rm sd}}{10^{38}\rm{erg/s}}\right)^{1/2},
\end{equation}
where $r_{\rm LC} = 1.6\times 10^8\rm{cm}$ is the light cylinder radius of the Crab Pulsar.
The radius of the termination shock of Crab Nebula is $\sim 10^9 r_{\rm LC}$ and 
spindown luminosity of the Crab Pulsar is $\sim 6\times 10^{38}\rm{erg/s}$.
Thus, the strength parameter of the entropy mode is estimated as $a \simeq 80$ at the termination shock.
Since the EM wave is expected to be converted from this entropy wave, 
we can expect this wave to be SuperLuminal "Strong" Wave (SLSW).

Next, we compare the inertial length and the wavelength of the entropy mode in upstream and downstream 
by assuming that the typical length scale does not change even if some fraction of energy is converted 
from the EM energy to the kinetic energy, such as the magnetic reconnection 
(Lyubarsky \& Kirk 2001).
If the wavelength of the entropy mode is shorter than the inertial length, the MHD approximation breaks down
and entropy wave can be converted to the other waves.
The ratio of these length scales at the upstream is 
\begin{equation}
\label{upratio}
\frac{\Gamma \lambda_{\rm sw}}{2\pi c/\omega_{\rm p,up}} \equiv \eta_{\rm up} \sim 2.7\times 10^2 \sqrt{n_{\rm up}\Gamma^2},
\end{equation}
where $\Gamma$ is the bulk Lorentz factor of the pulsar wind, $\lambda_{\rm sw}= 10^9\rm{cm}$ is the wavelength 
of the striped wind in the observer frame, $\omega_{\rm p,up}= \sqrt{4\pi n_{\rm up}e^2/m}$ is the upstream plasma frequency,
and $n_{\rm up}$ is the comoving number density.
Here we assumed that the plasma does not have relativistic temperature.
On the other hand, constraint on $n_{\rm up}$ and $\Gamma$ at the termination shock is obtained
from the spindown luminosity.
By expressing the spindown luminosity as a sum of isotropic kinetic and Poyniting fluxes,
and substituting the radius of the termination shock of the Crab Nebula into it, we obtain
\begin{equation}
\Gamma^2(1+\sigma)n_{\rm up} = 2\times 10^{-2}.
\label{wcon}
\end{equation}
Here $\sigma $ is the ratio of the Poynting flux to the kinetic flux.
Using this constraint, equation (\ref{upratio}) gives
\begin{equation}
\eta_{\rm up} \sim 3.8\times 10^1 \times (1 + \sigma)^{-1/2}.
\end{equation}
We can estimate $\eta_{\rm up} $ at the immediate upstream of the termination shock by using $\sigma$.
At the light cylinder, the pair cascade models predict $\sigma \sim 10^4$ (e.g. Hirotani 2006).
It should be reduced by the magnetic reconnection in the wind region.
If this process is extremely inefficient and $\sigma$ is close to $10^4$ at the termination shock, 
$\eta_{\rm up}<1$ is realized.
In this case, the entropy wave may break down by non-MHD effect.
On the other hand, if we adopt more a conventional value of $\sigma < 10^3$,  $\eta_{\rm up}$ should be larger than unity.
In this case, the entropy wave can survive in upstream.
Hereafter we consider the latter case.

The situation drastically changes in the downstream.
First we estimate the ratio in the downstream $\eta_{\rm down}$ under the assumption that the entropy mode does not convert to the SLSWs.
From this estimation, we can realize that this assumption is not appropriate.
Then, we consider an alternative scenario.

According to the observational fact, the bulk Lorentz factor of the downstream plasma is $\Gamma \sim1$.
The wavelength of the entropy wave changes due to shock compression, but their compression ratio measured in the downstream frame is at most $O(1)$.
Therefore, this scale is $\sim \lambda_{\rm sw}$ in the downstream frame. 
On the other hand, the inertial length in the downstream frame increases drastically because of the thermalization.
The ratio of them becomes
\begin{equation}
\eta_{\rm down} =\frac{\lambda_{\rm sw}}{2\pi c/\omega_{\rm p,down}} = (\gamma_{\rm th}\Gamma)^{-1/2}\eta_{\rm up},
\end{equation}
where $\gamma_{\rm th}$ is the Lorentz factor of the thermal motion of the downstream plasma,
$\omega_{\rm p,down}= \sqrt{4\pi n_{\rm down}e^2/\gamma_{\rm th}m}$ is the plasma frequency which takes into account the
relativistic correction, and $n_{\rm down}$ is the downstream number density.
We assume $\gamma_{\rm th}\sim \Gamma$, since the downstream bulk velocity is nonrelativistic.
As a result, $\eta_{\rm down}$ is smaller than $\eta_{\rm up}$ by a factor of $\Gamma$.
This bulk Lorentz factor is estimated as $10^2\leq\Gamma\leq10^6$ by pair cascade models 
(e.g. Hirotani 2006),
many MHD models of Pulsar Wind Nebula (PWN), and the induced Compton scattering constraint (Tanaka \& Takahara 2013).
Using the constraint, we estimate $\eta_{\rm down}$ to be
\begin{equation}
10^{-6}\lesssim \eta_{\rm down} \lesssim 1.
\end{equation}
Thus, MHD approximation is not adequate to describe the entropy waves in the downstream,
and it should be dissipated (Petri \& Lyubarsky 2007, Sironi \& Spitkovsky 2011)
or converted to some other waves (Amano \& Kirk 2013).

Before proceeding further, let us summarize the above estimates.
The strength parameter $a$ of the SLSWs is estimated as $O(10)$.
The wavelength of the entropy mode is expected to be longer than the inertial length in upstream,
while the opposite is expected in the downstream.
The expected values of their ratios are $\eta_{\rm up} \gtrsim 1$ and $\eta_{\rm down}\sim 10^{-3}$.

The entropy mode cannot be sustained by the downstream plasma.
It should be converted to EM waves by the interaction with the shock front and the EM waves propagate 
back to the upstream region (Amano \& Kirk 2013).
As we will see below, the entropy mode can convert to SLSWs even in the upstream.
Hereafter we consider the propagation condition of a SLSW, which is affected by the
strength of the EM wave and the thermalization of the background plasma.

First, we review the effect of the strength of the wave.
The frequency of the SLSW should be $\sim 2\pi c/\lambda_{\rm sw}$ in the shock rest frame.
Here, we assume that this SLSW propagates toward the shock front in this frame.
The frequency of this SLSW in the upstream frame is $2\pi c/\Gamma\lambda_{\rm sw}$.
For an ordinary ($a<1$) EM wave in the plasma, the condition for the propagation is 
$\omega^2>\omega_{\rm p}^2$.
This condition coincides with $\eta_{\rm up}<1$.
However, for SLSW, the condition is modified by the strength parameter.
The dispersion relation of the SLSW for pair plasma is described by Kaw \& Dawson (1970) as 
\begin{equation}
\omega^2 = \frac{2\omega_{\rm p}^2}{\sqrt{1+a^2}}+k^2c^2,
\label{dr}
\end{equation}
where $k$ is the wavevector.
Hence, the condition for the propagation becomes 
\begin{equation}
\frac{\omega^2}{\omega_{\rm p}^2}>\frac{2}{\sqrt{1+a^2}}.
\end{equation}
Therefore, the wave can propagate when $a$ is sufficiently large even if $\eta_{\rm up}>1$.

Next, we discuss the thermalization.
The SLSWs in the over dense plasma (i.e. $\eta>1$) are unstable (Max \& Parkins 1971),
and also it is shown by the two-fluid simulation that these waves generate EM waves and sound waves 
by the stimulated Brillouin scattering (Amano \& Kirk 2013).
As a result, the SLSWs thermalize the plasma in the upstream region by the dissipation of sound waves. 
If this mechanism works well, the upstream plasma gain a relativistic thermal energy.
The plasma frequency becomes lower as Lorentz factor of the thermal motion becomes large as 
$\omega_{\rm p}\propto \gamma_{\rm th}^{-1/2}$.
Thanks to this nonlinear effect, the local plasma frequency in upstream becomes lower than the SLSW frequency.
The instabilities of the SLSWs are not fully understood.
The growth rate is only known for limited conditions (c.f. Max \& Parkins 1972, Asseo et al. 1978, Lee \& Lerche 1978).
It tends to be low when $a$ is not much larger than $1$ and $\omega$ is sufficiently larger than $\omega_{\rm p}$
 (Amano 2014).
This is natural because the wave becomes an ordinary EM wave in vacuum in the limit of $a\ll1$ and $\omega\gg\omega_{\rm p}$.

As we have seen above, SLSWs can exist in the upstream region of the termination shock.
In this paper, we study the electron acceleration in the upstream rest frame.
We assume that $a= O(10)$ and $\omega>\omega_{\rm p}$.
These assumptions are acceptable near the immediate upstream region.

\subsection{Setup}
We describe the EM turbulences by the superposition of the EM waves.
In this paper, we ignore wave-wave interactions for simplicity.
Such interaction may not be negligible and will be treated in future works.
This description of the turbulence is based on Giacalone \& Jokipii (1999), 
in which they calculated the transport of the cosmic rays in the static magnetic field.
Recently, they studied particle accelerations using this scheme (e.g., Giacalone \& Jokipii 2009, Guo \& Giacalone 2014).
Here, we note the difference between our assumption on the waves and theirs.
We assume propagating (superluminal) EM waves.
On the other hand, their magnetic field is static in the fluid rest frame (entropy waves), 
and the motional electric field
$\vec{E} = -(\vec{v}\times\vec{B})/c$ is used to calculate the particle acceleration in the shock rest frame,
where $\vec{v}$ and $\vec{B}$ are the fluid velocity and magnetic field, respectively.
This difference has a big impact on the particle acceleration mechanisms such as the phase locking.

The electric field and magnetic field are expressed by the superposition of elliptically-polarized waves.
They are decomposed to linearly polarized sinusoidal waves.
Strictly speaking, the wave form of SLSW does not have a pure sinusoidal shape, 
but rather a sawtooth like shape (Max \& Parkins 1971).
Our approximation of sinusoidal wave is adequate for $\omega\gg\omega_{\rm p}$.
We assume that there are a primary wave which is generated from entropy mode and isotropically distributed daughter EM waves.
It is assumed that the entropy mode is completely transformed to the primary wave for simplicity.
The primary wave is assumed to be linearly polarized and propagates to $z$-direction as 
\begin{eqnarray}
\vec{E}_0 & = &  A_0 \cos(\omega_0 t - k_0 z)\hat{e}_x, \\
\vec{B}_0 & = & (A_0/\beta_{{\rm ph},0})\cos(\omega_0 t - k_0 z)\hat{e}_y,
\end{eqnarray}
where $A_0$, $\omega_0$, $k_0$, $\beta_{{\rm ph},0}$ are the amplitude, frequency, wavenumber and phase velocity, respectively.
Here $\hat{e}_x$ and $\hat{e}_y$ are the unit vector toward $x$-direction and $y$-direction, respectively.
The secondary components are described as 
\begin{eqnarray}
\vec{E}_{\rm sec}(\vec{x},t) &=& \sum_{n=1}^N A_n\exp{\{i(\vec{k}_n\cdot\vec{x} -\omega_n t + \zeta_n)\}}\hat\xi_{E,n} \\
\vec{B}_{\rm sec}(\vec{x},t) &=& \sum_{n=1}^N \frac{A_n}{\beta_{\rm ph,n}}\exp{\{i(\vec{k}_n\cdot\vec{x} -\omega_n t+ \zeta_n)\}}\hat\xi_{B,n},
\end{eqnarray}
where $A_n$, $\vec{k}_n$, $\omega_n$, $\zeta_n$ are the amplitude, wavevector, frequency, and phase of each mode, respectively.
Here $\beta_{{\rm ph},n} = v_{\rm ph,n}/c$ is the phase velocity of each mode, which is calculated from the 
dispersion relation (equation (\ref{dr})) as we will show later.
Since the phase velocities are larger than unity, the amplitude of the electric field is larger than the magnetic field.
We note that it can be understood by considering the Faraday's law 
$\vec{\nabla}\times \vec{E} = -\frac{1}{c}\frac{\partial\vec{B}}{\partial t}$.
Each waves propagate toward $\hat{e'}_{z,n} = \vec{k}_n/|k_n|$, 
with which $\hat{e'}_{x,n}$ and $\hat{e'}_{y,n}$ form the orthogonal coordinate system.
The polarization vector are written as $\hat{\xi}_{E,n} = \cos \psi_n \hat{e'}_{x,n}+i\sin\psi_n\hat{e'}_{y,n}$ and 
$\hat{\xi}_{B,n} = -i\sin \psi_n \hat{e'}_{x,n}+\cos\psi_n\hat{e'}_{y,n}$.
Since the distribution of the secondary waves is uncertain, we assume it to be isotropic.
In this paper, $e'_{z,n}$ is chosen randomly to make the distribution of the secondary waves isotropic.
The polarization ($\hat{\xi}_{E,n}$ and $\hat{\xi}_{B,n}$) and phase $\zeta_n$ are also randomly distributed.
The amplitude of each mode is given by
\begin{equation}
A_n^2 = \varsigma_{\rm sec}^2G_n\left[\sum_{n=1}^NG_n\right]^{-1},
\end{equation}
where $\varsigma_{\rm sec}^2$ represents the mean intensity of the secondary waves.
The number of Fourier components $N$ is $10^2$ in this paper.
We use the following form for the power spectrum
\begin{equation}
G_n = \frac{4\pi\omega_n^2\Delta\omega_n}{1+(\omega_nT_c)^\alpha},
\end{equation}
where $T_{\rm c}$ is the coherence time which is set as $\omega_0T_{\rm c} = 1$,
and $\alpha-2$ is the power law index of the energy spectrum of the secondary waves.
We set $\alpha = 11/3$, which makes Kolmogorov-like turbulences. 
Here, $\Delta \omega_n$ is chosen such that there is an equal spacing in logarithmic $\omega$-space,
over the finite interval $\omega_{\rm min}\leq\omega\leq\omega_{\rm max}$.
We set the minimum frequency and the maximum frequency as $\omega_{\rm min} = \omega_0$
and $\omega_{\rm max} = 10^3\omega_0$ in all calculations.
The sum of the EM energy density of the primary and secondary waves is set to be constant in each run as
\begin{equation}
\varsigma^2 = A_0^2 + \varsigma_{\rm sec}^2.
\end{equation}
Using $\varsigma$, the unit of time in this paper is defined as
\begin{equation}
\frac{mc}{e\varsigma} \equiv 1
\end{equation}
The strength parameter is defined by using $\varsigma$ and $\omega_0$ as
\begin{equation}
a \equiv \frac{e\varsigma}{mc\omega_0}.
\end{equation}
When there is no primary wave, the strength parameter is defined by replacing $\omega_0$ to $\omega_{\rm min}$.
The phase velocities $\beta_{\rm ph,n}=\omega_{\rm n}/(k_n c)$ are calculated from the dispersion relation   
\begin{equation}
\omega^2_n = \frac{2\omega_{\rm p}^2}{\sqrt{1+a^2}} + k_n^2c^2.
\end{equation}
Here we note that the superposition of the modes is an approximation,
because this dispersion relation is nonlinear.
The appropriateness of this approximation for the obtained results should be checked in future works.
We assume that the electrons initially have the relativistic energy $\gamma_0mc^2 = 10mc^2$
and have isotropic velocity distribution which mimics thermalized particles in the precursor region of the termination shock.
In this paper we use $10^4$ electrons for the calculation.
Since we neglect the back reaction from particles to EM fields, the electron number is important 
only for the statistics and does not change the physics.
The injection points of these particles are chosen to be homogeneous.
We fix the plasma frequency as $\omega_{\rm p} = \sqrt{5}\times 10^{-2}\omega_0$
and it does not change in time.
This fulfills the stable propagation condition.
It is lower than the plasma frequency which is estimated from the initial thermal Lorentz factor.
However, as the Lorentz factor of the electrons becomes higher, the plasma frequency becomes lower.
To avoid complexity, we approximate the plasma frequency by a fixed value, which corresponds to $\gamma_{\rm th} \sim 100$.

We inject relativistic electrons as test particles in this prescribed EM field.
We solve the equation of motion 
\begin{equation}
\frac{d}{dt}(\gamma m_{\rm e}\vec{v})=-e(\vec{E} + \frac{\vec{v}}{c}\times \vec{B})
\end{equation}
by using the Buneman-Borris method.
We neglect the radiation back reaction. This is an adequate approximation for the
electrons which have not been injected in shock crossing process,
because the cooling timescale is much longer than the dynamical timescale (Amano \& Kirk 2013).
The radiation spectra are calculated from the information of the motion by direct using the Lienard-\textbf{Wiechert} potential  
\begin{equation} 
  \frac{dW}{d\omega d\Omega} = \frac{e^2}{4 \pi c^2} 
  \Bigl| \int^{\infty}_{-\infty} \:dt^{\prime} \frac{ \vec{n} \times \bigl[ (\vec{n} - \vec{\beta}) \times \dot{\vec{\beta}} \bigr] } 
  {(1 - \vec{\beta} \cdot \vec{n} )^2 }\exp\bigl\{{i\omega ( t^{\prime} - \frac{\vec{n} \cdot \vec{r}(t^{\prime})}{c})}\bigr\} \Bigr|^2,
\label{L-W}
\end{equation}
where $\vec{\beta}=\vec{v}/c$ is the velocity of the electron, $\vec{n}$ is the observer direction, and $t'$ is the retarded time 
(Hededal 2005).
This formula can be applied for the frequency range of $\omega >\gamma \omega_{\rm p}$ for the radiation in the plasma.
In the next section, we will calculate the radiation spectra.
We will see that the above condition satisfied almost all frequency range.
When the exception is encountered, we will note it.

\section{Results}
\label{results}
\subsection{Particle acceleration}
\begin{figure}
\includegraphics[width = 16cm,angle=0]{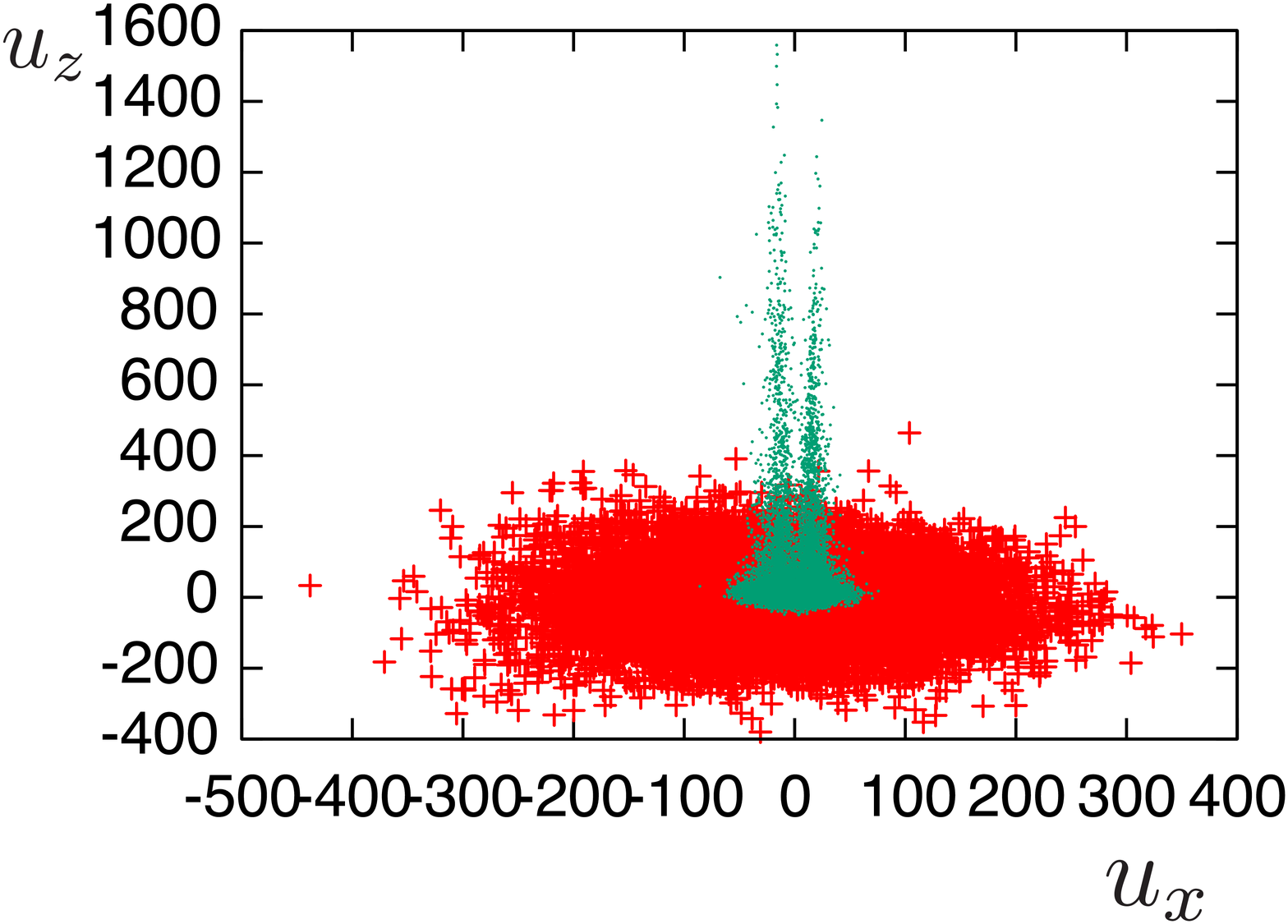}
\caption{Distribution of 4-velocity for $e\varsigma_{\rm sec}/mc = 0.1$ (green) 
and $1$ (red) at $t = 3\times 10^4\omega_0^{-1}$.
}
\label{4-velocity}
\end{figure}

First, we demonstrate one of the notable features of the particle acceleration by the SLSW,
namely the strong acceleration toward the wavevector direction.
For this calculation, we set $\omega_0 = 0.1$.
Since the unit of frequency is $e\varsigma/mc$ as noticed earlier, the strength parameter is $a=e\varsigma/mc\omega_0 = 10$.
We show the distribution of the $x$ and $z$ components of a $4$-velocity at $t = 3\times 10^4\omega_0^{-1}$ in Fig \ref{4-velocity}.
The red cross' are the 4-velocities for $e\varsigma_{\rm sec}/mc = 1$, which means that the turbulence consists of
the secondary waves without primary wave.
On the other hand, green dots are the $4$-velocities for $e\varsigma_{\rm sec}/mc = 0.1$.
The corresponding amplitude of the primary wave satisfies $eE_0/mc = e\sqrt{(\varsigma^2 - \varsigma_{\rm sec}^2)}/mc \simeq 0.995$, and
therefore the primary wave is dominant in this case.
The red cross' distribution is nearly isotropic, while the green dots' distribution is quite anisotropic.
The anisotropy is due to the acceleration of the electrons toward the wavevector of the primary wave.
Even though there are other waves, the primary wave can dominate the electron motion 
and the phase locking occurs for $e\varsigma_{\rm sec}/mc = 0.1$.
The distribution for $x$-direction is symmetric and the width is around $10$ times smaller than 
that for the case with $e\varsigma_{\rm sec}/mc = 1$.
This is because the amplitude of the secondary components $\varsigma_{\rm sec}$ is $10$ times smaller.

\begin{figure}
\includegraphics[width = 16cm]{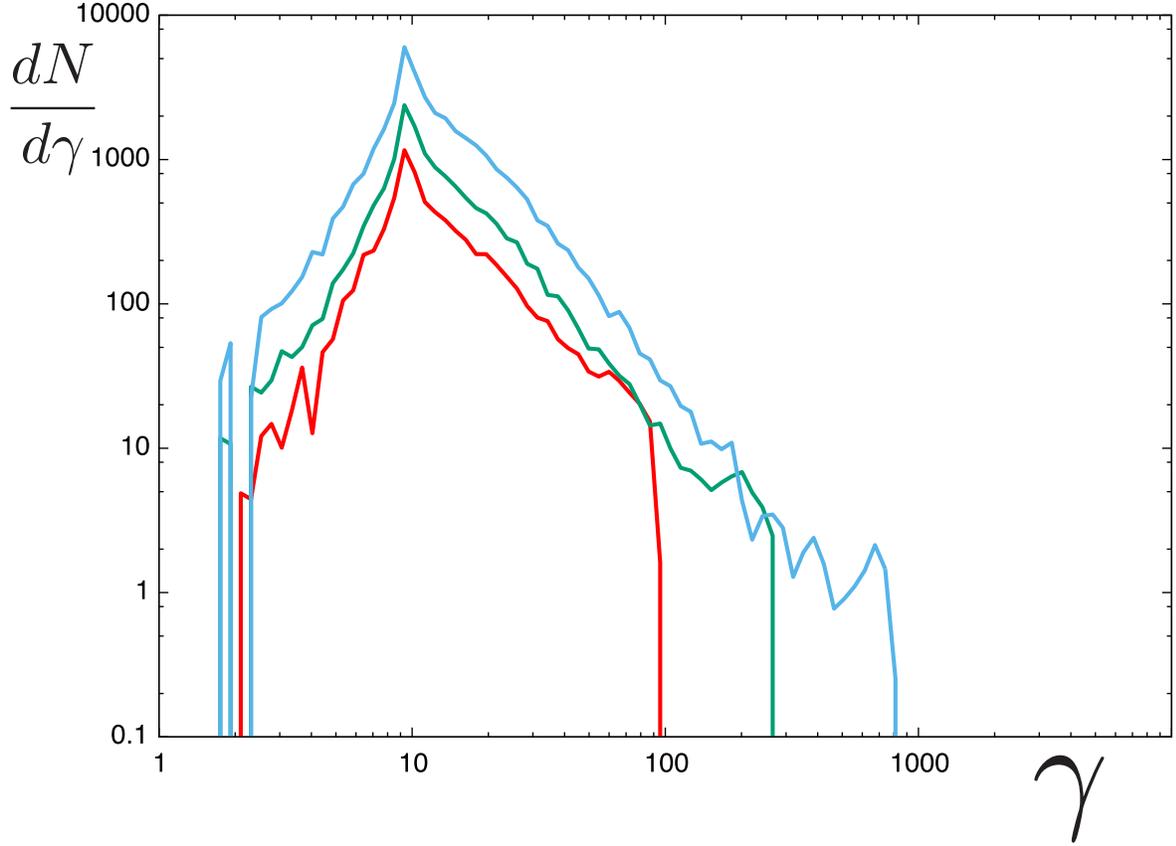}
\caption{The evolution of the energy spectrum for $e\varsigma_{\rm sec}/mc = 10^{-3}$.
Each lines are the spectra for $t = 30\omega_0^{-1}$ (red), $300\omega_0^{-1}$ (green) and $3000\omega_0^{-1}$ (light blue).
The green and light blue lines are shifted in vertical direction by a factor of $2$ and $5$, respectively,  
to see their difference clearly.
}
\label{ene-spe-evo}
\end{figure}

In Fig \ref{ene-spe}, we show the energy spectra for different amplitudes of secondary components 
($e\varsigma_{\rm sec}/mc = 1,\ 0.5,\ 0.1,$ and $10^{-3}$) at $t = 3\times10^4\omega_0^{-1}$.
For $e\varsigma_{\rm sec}/mc = 10^{-3}$, the energy distribution has a pure power law distribution.
In this case, the secondary components are quite weak compared to the primary component.
As a result, electrons which are initially moving nearly parallel to the wavevector of the primary wave are 
in the state of "phase locking" and selectively accelerated.
The cutoff energy reaches an expected value of $a^2\gamma_0 = 10^3$ in the timescale of 
$a^2\gamma_0^2\omega_0^{-1} = 10^4\omega_0^{-1}$, which is the typical oscillation timescale of a "phase locked" particle.
The other electrons also tend to accelerate to the wavevector direction of the primary wave,
but this acceleration is weaker, because they are not in the phase locking state.
To clarify the selective acceleration, we show the time series of the energy spectra 
for $e\varsigma_{\rm sec}/mc = 10^{-3}$ in Fig \ref{ene-spe-evo}.
The high energy cutoff evolves in time due to the selective acceleration.
On the other hand, the peak Lorentz factor does not vary and remains at $\gamma\sim 10$, 
since the acceleration by secondary waves is quite weak in this case.
This distribution shows a power law shape of $dN/d\gamma \propto \gamma^{-2}$.
We do not intend to claim that this power law index is universal, because it can be altered by initial conditions.
For example, the initial velocity (direction) distribution, assumed to be isotropic, apparently changes the
resultant energy distribution, since the phase locking effect is strongly dependent on the angle between the velocity and wavevector of the SLSW.
Here we stress that we obtain high energy electrons which have an energy much higher than the peak energy,
and the phase locking effect by the strong wave is an important factor to realize it.

For $e\varsigma_{\rm sec}/mc = 1$,
many EM waves accelerate the electrons without long term phase locking, which is prevented by the disturbance 
of the velocity direction by the other EM waves.
As a result, electrons diffuse in the momentum space, and the distribution function $f(p)$ is well described by the Gaussian function.
We can see that the energy distribution $dN/d\gamma\propto p^2f(p)$ is consistent with the power law distribution 
in the low energy side with the index $+2$ 
and an exponential cutoff in the high energy side.

Lastly let us focus on the cases for $e\varsigma_{\rm sec}/mc = 0.1$ and $0.5$.
In these cases, both the primary and secondary waves affect the particle distribution 
(green and light blue line in Fig \ref{ene-spe}).
While most particles are diffusively accelerated by the turbulence (secondary waves), 
a small fraction of particles are selectively accelerated by the primary wave.
This selection can clearly be seen in $4$-velocity distribution for $e\varsigma_{\rm sec}/mc = 0.1$ in Fig \ref{4-velocity}.
The peak energy difference between two distributions comes from the energy density of the secondary waves.
The electrons in this peak energy range are accelerated diffusively by the secondary waves.
Therefore, the energy density of the secondary waves is smaller and the peak energy is lower.
If the $A_0$ is slightly larger than $\varsigma_{\rm sec}$ in the upstream of the termination shock,
this acceleration mechanism can produce a broad energy distribution which contains a high energy power law tail.

\subsection{Radiation}
\begin{figure}
\includegraphics[width = 16cm]{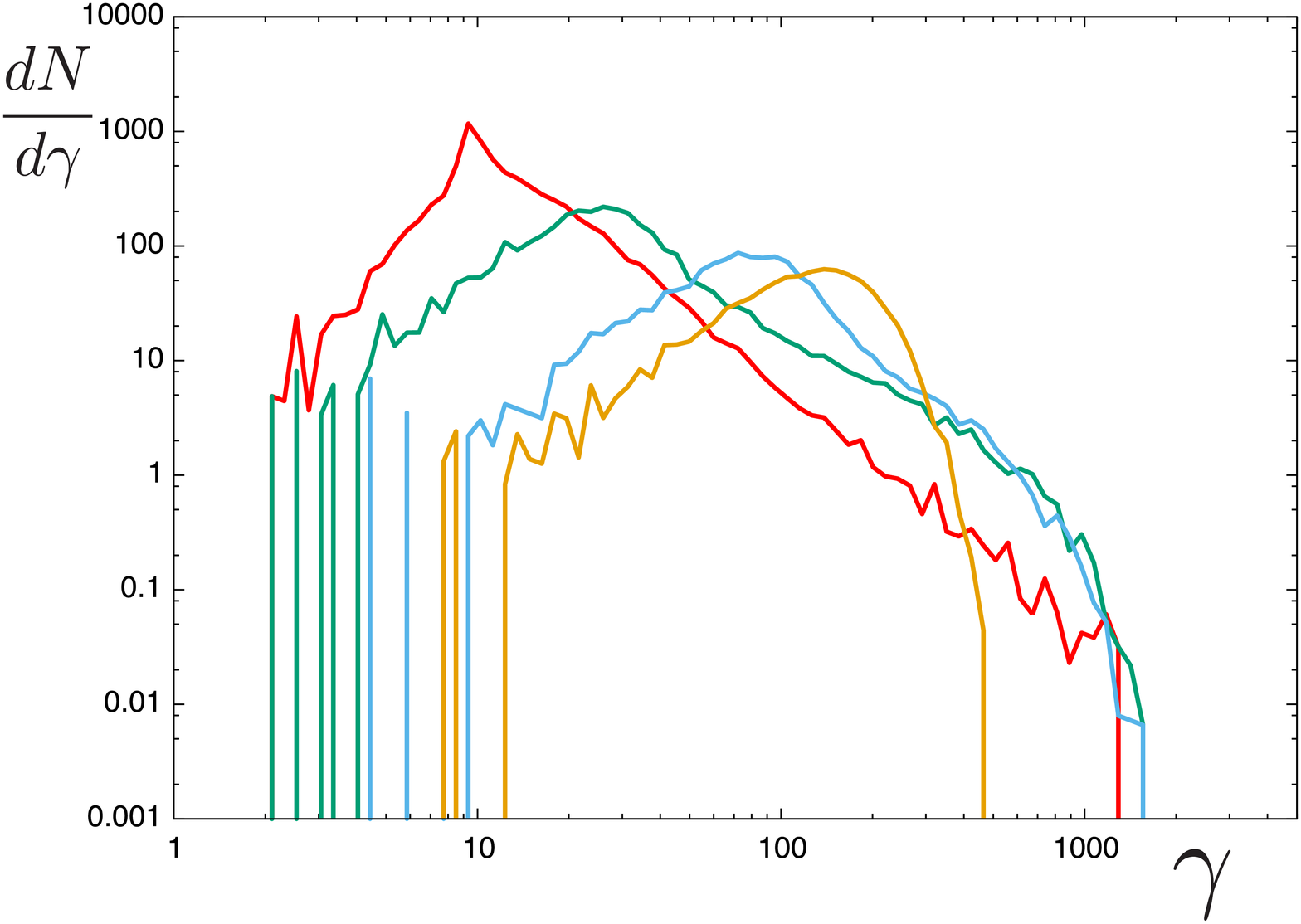}
\caption{Particle energy distributions for $t = 3\times10^4\omega_0^{-1}$ and $\alpha = 11/3$.
The horizontal axis is Lorentz factor of the electrons, and the vertical axis is $dN/d\gamma$.
The curved lines are corresponding to $e\varsigma_{\rm sec}/mc=10^{-3}$ (red), $0.1$ (green), $0.5$ (light blue) and $1$ (yellow).
}
\label{ene-spe}
\end{figure}

\begin{figure}
\includegraphics[width = 16cm]{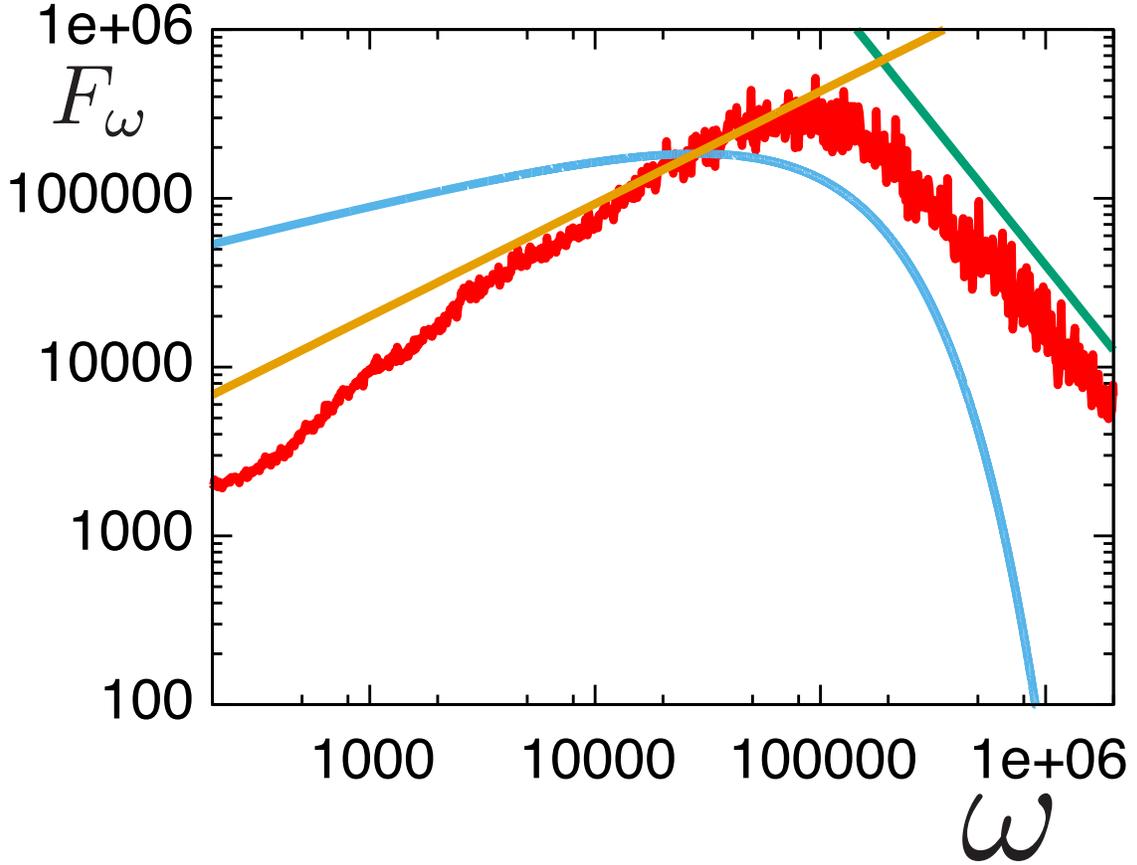}
\caption{Radiation spectra from the accelerated electrons for $e\varsigma_{\rm sec}/mc = 0.1$ at $t= 3\times 10^4\omega_0^{-1}$.
Vertical axis is the flux in arbitrary unit and horizontal axis is frequency in unit at $e\varsigma/mc$.
The observer direction is $\vec{n} = (10^{-2},0,\sqrt{1-10^{-4}})$, 
which is nearly parallel to the wavevector of the primary wave $\hat{e}_z$.
The curved light blue line is the synchrotron theoretical curve.
The straight green line is $F_\omega \propto \omega^{-5/3}$ and straight brown line is 
$F_\omega \propto \omega^{2/3} $ for comparison.
}
\label{rad-spe-nx-2}
\end{figure}

\begin{figure}
\includegraphics[width = 16cm]{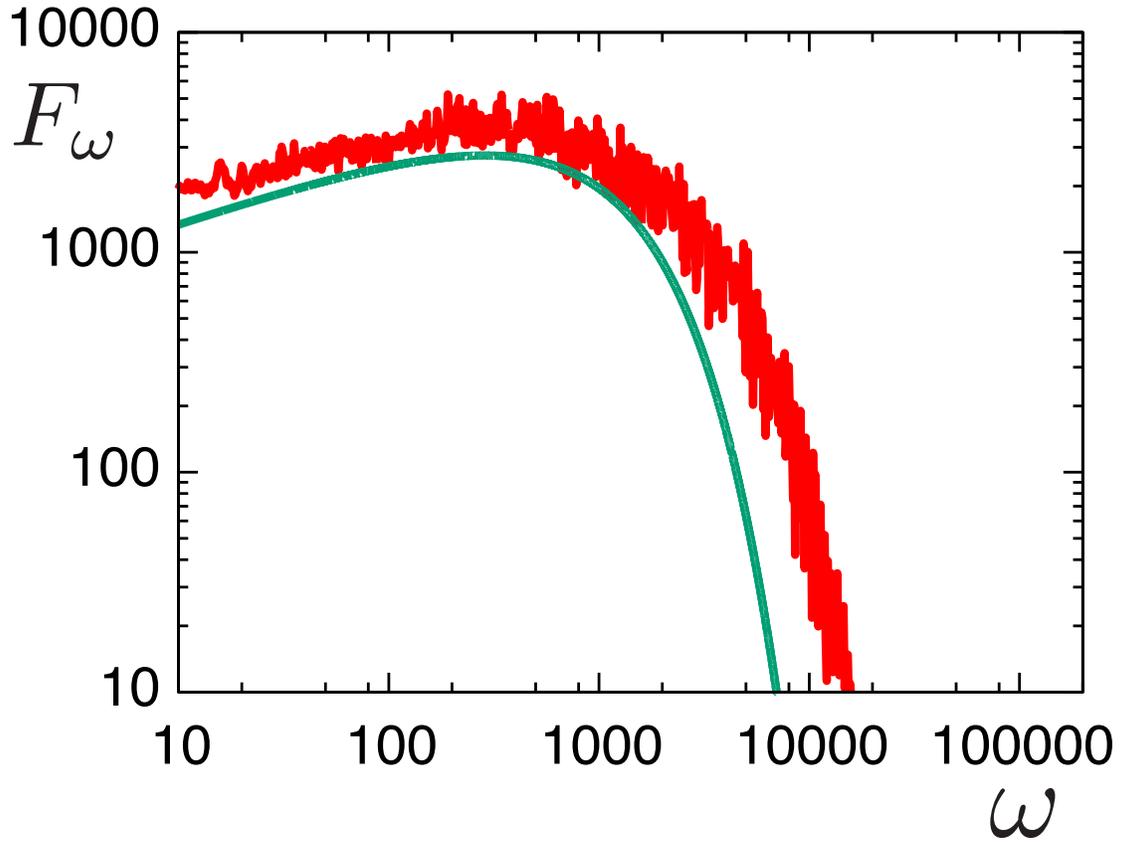}
\caption{Radiation spectra from the accelerated electrons for $e\varsigma_{\rm sec}/mc = 0.1$ at $t= 3\times 10^4\omega_0^{-1}$.
The observer direction is in the $x$ direction.  
The curved (green) line is the synchrotron theoretical curve for comparison.
}
\label{rad-spe-nx1}
\end{figure}

The radiation spectra of the electrons moving in the SLSWs
for $e\varsigma_{\rm sec}/mc = 0.1$ are shown in Figs \ref{rad-spe-nx-2} and \ref{rad-spe-nx1}.
We calculate the radiation spectra in the restricted time from $t = 3\times 10^4\omega_0^{-1}$ to 
$t = (3\times 10^4 + 2\times 10^3)\omega_0^{-1}$.
The starting time $t = 3\times 10^4\omega_0^{-1}$ corresponds to the time of the energy distribution of Fig \ref{ene-spe}.
This integration timescale is longer than Photon Formation Time 
of the typical frequency of the synchro-Compton radiation by a factor of $ 2\times10^3$.
This ensures that we can resolve the radiation spectrum down to $\sim 10^{3}$ times lower frequencies than the peak one.
On the other hand, the time step for pursuing the electron motion is $10^{-2}$.
This ensures that we can resolve the radiation spectrum up to $\sim 10^2$ times higher frequencies than the peak one.
Strictly speaking, the time step should be smaller than the inverse of the radiation frequency, 
and it is much shorter than $10^{-2}$.
However, it is showed that the radiation spectrum is well described by using the large time step which is about only one order 
shorter than the photon formation time (Reville \& Kirk 2010).
The horizontal axis is the frequency normalized by $e\varsigma/mc$.
The vertical axis is the flux in an arbitrary unit.
The jaggy lines are the calculated radiation spectra, and the smooth lines are the analytical synchrotron curve,
which are shown for comparison.

In Fig \ref{rad-spe-nx-2}, the observer direction is nearly along the $z$ direction. To be precise, we set the observer direction 
$\vec{n} = (n_x,n_y,n_z) = (10^{-2},0, \sqrt{1-10^{-4}})$.
The reason for not choosing $\vec{n} = (0,0,1)$ will be explained later.
The peak frequency of the radiation spectrum in Fig \ref{rad-spe-nx-2} is $\omega_{\rm peak} \sim 10^5$.
This can be understood by using the synchro-Compton theory (Rees 1971, Gunn \& Ostriker 1971).
The cutoff Lorentz factor $\gamma_{\rm cut}$ is around $10^3$ for $e\varsigma_{\rm sec}/mc = 0.1$ 
at $t = 3\times 10^4\omega_0^{-1}$ as is seen in Fig \ref{ene-spe}.
The peak frequency can be estimated as 
$\omega_{\rm peak} \sim \gamma^2 e\varsigma_{\rm sec}/mc = 10^5$,
in the same manner as the synchrotron radiation. 
This can be justified as follows.
First, we consider an electron which is in the phase locking state in a strong wave and tentatively ignore the other waves.
In this case, the Lorentz force is negligible except for the direction parallel to the velocity.
The resultant trajectory is nearly straight, 
and the curvature radius of the orbit is very long compared to $\gamma mc^2/e\varsigma_{\rm sec}$.
Next we add the other waves which are isotropically distributed.
The electron trajectory obtains a wiggling shape with the curvature radius for each case to be $\sim \gamma mc^2/e\varsigma_{\rm sec}$.
The deflection angle during a typical deflection is $\sim 1/\gamma$,
since the strength parameter defined by using only secondary components such as $e\varsigma_{\rm sec}/mc\omega_{\rm min}$ is unity.
As a result, electrons emit the "synchro-Compton" radiation around the peak frequency.
The large part of this radiation power comes from the electrons moving nearly along the $z$-direction,
since the selectively accelerated electrons are moving in this way.
Thus, the radiation spectrum has a clear peak at $\omega \sim \gamma_{\rm max}^2e\varsigma_{\rm sec}/mc\sim 10^5$.
We note that the deviation from the exponential cutoff in the highest frequency region comes from the jitter radiation contribution,
since there are secondary waves which have higher frequency than $\omega_{\rm min} = e\varsigma_{\rm sec}/mc$ (cf. Teraki \& Takahara 2011).
The spectrum in the frequency region lower than the peak is harder than the ``isotropic" synchrotron theoretical one.
In our case, the velocity distribution is quite anisotropic. 
The spectral index of the synchrotron radiation toward some direction, due to an electron, is $2/3$ (cf. Jackson 1999).
The calculated spectrum seems to be slightly harder than $F_\omega \propto \omega^{2/3}$,
so that additional mechanisms may contribute to this spectrum, 
but we do not discuss this topic further.
More detailed analyses will be done in our future work.
The important thing which we should stress here is that 
the primary wave does not contribute to the radiation directly.
It works only for the energy gain of the electrons.
The radiation power and typical frequency are determined by the energy of the electrons 
and EM energy density of the secondary waves.

In Fig \ref{rad-spe-nx1}, we set $\vec{n} = (1,0,0)$.
The spectrum is well described by the isotropic synchro-Compton radiation.
The spectral index at the frequency region lower than the peak coincides with $1/3$,
and the peak frequency $\omega_{\rm peak} \sim 10^3$ is the expected value. 
The Lorentz factor of the electrons moving around the $x$-direction is $\gamma\sim 30$ 
(see Fig \ref{4-velocity} and Fig \ref{ene-spe}).
The magnetic field ($y$-direction) of the primary wave mainly contributes to the radiation power for this case.
Thus, the peak frequency is described as $\gamma^2 eB_{0y}/mc \sim (30)^2\times1\sim 10^3$.
The spectrum at the frequency region higher than the peak shows slower decline than the exponential cutoff.
This is due to the superposition of the contributions from electrons with higher energies.
Jitter radiation components do not stand out in this spectrum, because of $eB_{0y}/mc\omega_0 \gg1$ (cf. Teraki \& Takahara 2011).

\begin{figure}
\includegraphics[width = 16cm]{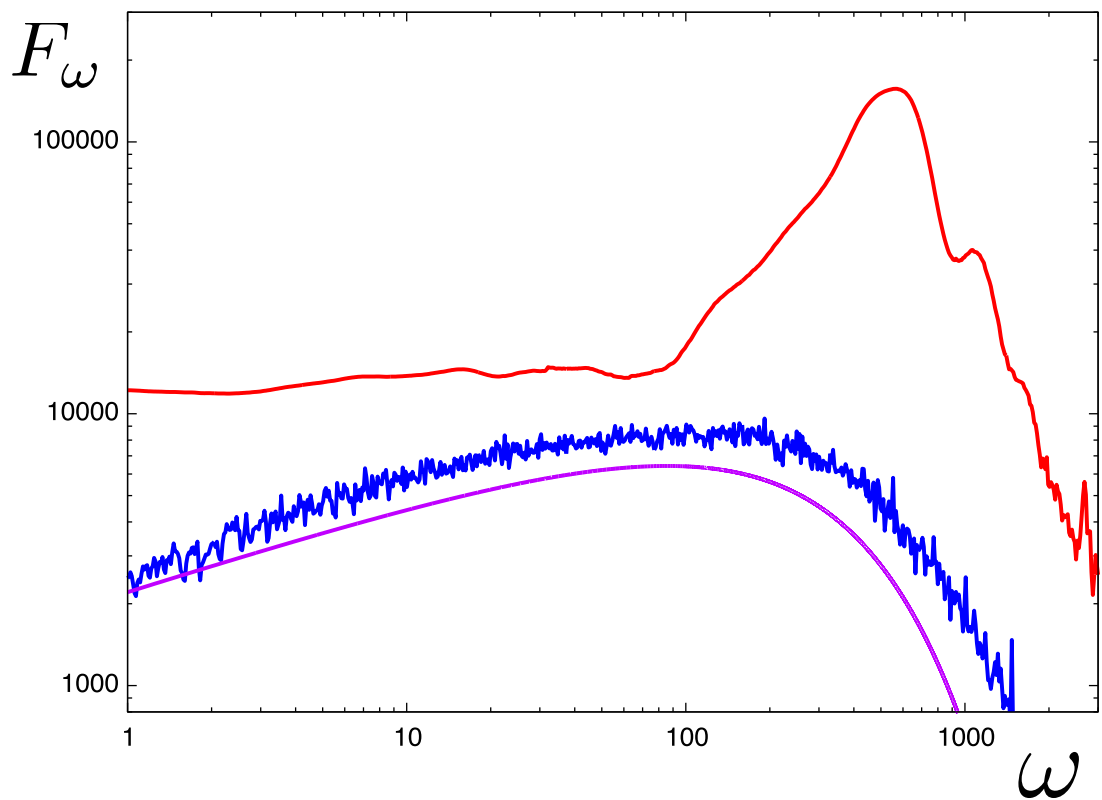}
\caption{Radiation spectra from the accelerated electrons for $e\varsigma_{\rm sec}/mc = 10^{-3}$ at $t = 3\times 10^4\omega_0^{-1}$.
The observer direction for upper (red)line is $\vec{n} = (10^{-2},0,\sqrt{1-10^{-4}})$.
The observer for the lower (blue) jaggy line is in the $x$ direction.
The curved (magenta) line near the upper line is the synchrotron theoretical curve.
}
\label{rad-spe-ss1d0}
\end{figure}

To confirm the fact that the scatterers needed for the radiation of Fig \ref{rad-spe-nx-2} are the secondary components,
we show the radiation spectra for $e\varsigma_{\rm sec}/mc = 10^{-3}$ in Fig \ref{rad-spe-ss1d0}.
The secondary waves are extremely weak compared to the primary wave, 
so that the primary wave causes the radiation.
The integration time is identical to that used for depicting Figs \ref{rad-spe-nx-2} and \ref{rad-spe-nx1}.
The lower (blue) line is the radiation spectrum for the observer located in the $x$-direction.
The peak frequency at $\sim 100$ can be explained in the same manner as above.
The $B_y$ component of the primary wave $B_{0y}$ strongly deflects the electron,
and produces the peak frequency at $\sim \gamma^2 eB_{0y}/mc\simeq100$,
since the typical Lorentz factor for the unselected electrons (which is not moving toward the $z$-direction)
is $\simeq 10$ and $eB_{0y}/mc \simeq 1$.
The spectral shape roughly coincides with the theoretical synchrotron curve as seen in Fig \ref{rad-spe-ss1d0}.
On the other hand, different features are found in the spectrum for the observer located in
$\vec{n} = (n_x,n_y,n_z) = (10^{-2},0, \sqrt{1-10^{-4}})$, which is shown by the red line.
It is noted that the spectrum is shifted vertically by a factor of $200$ to see the shape clearly.
The spectral shape is clearly different from the synchro-Compton radiation, 
because this radiation signature directly reflects the nonlinear orbit.
This large-scale orbit is determined by the strong primary wave, and the small-scale deflection angle
is much smaller than $1/\gamma$, since the strength parameter for the secondary waves 
$e\varsigma_{\rm sec}/mc\omega_{\rm min} = 10^{-2}$ is very small.
The sweeping behavior(of the beaming cone) is completely different from the sweeping 
with curvature radius $\gamma mc^2/eB$, which is assumed for the synchrotron radiation.
The particle trajectory is nearly straight, and the sweeping time is much longer than $mc/e\varsigma$ and $1/\omega_0$.
In this case, the sweeping timescale is roughly $10$ times shorter than 
the maximum phase locked oscillation timescale $2\pi a^2\gamma_0^2\omega_0^{-1}$.
In our calculation, the initial velocity directions do not coincide the wavevector of the primary wave,
and $|\vec{E}|\neq|\vec{B}|$.
Moreover, the phase velocity of this wave is larger than $c$.
Such effects shorten the sweeping timescale by a few times compared to the case for the vacuum.
As a result, the typical frequency is $\gamma_{\rm max}^2\omega_0/(\gamma_0^2a^2) \times O(10)\gtrsim 10^2$.
We note that the second harmonics can be seen around $\omega\sim 1000$.
We note that this radiation spectra in the lowest frequency region $\omega\simeq10$ is not precise
because $\gamma\omega_{\rm p} \sim 10^3\times \sqrt{5}\times10^{-3}$ is $ O(1)$.
However, this frequency range is not important for the current discussion.
Here, we showed that the scatterer for the radiation for $e\varsigma_{\rm sec}/mc=10^{-3}$ is the primary wave.
Furthermore, we confirmed that the radiation signature is not always described by the synchro-Compton theory.
From this fact, we can understand that the scatterers which realize the spectra in Fig \ref{rad-spe-nx-2} 
are the secondary waves.

Lastly let us explain the reason for choosing the observer direction to be 
$\vec{n} = (n_x,n_y,n_z) = (10^{-2},0, \sqrt{1-10^{-4}})$, and not $\vec{n} = (0,0,1)$.
As mentioned above, the selectively accelerated electrons tend to move toward the $z$-direction, but there are
very few electrons moving ''very close" to the $z$-axis.
The reason is as follows.
When the phase locking occurs, the perpendicular forces (to the velocity) from the electric magnetic fields 
nearly cancel each other.
From the balance of perpendicular forces, we can estimate the angle $\theta$ between the wavevector and the velocity.
By using the assumed parameter $\omega_{\rm p} = \sqrt{5}\times 10^{-2}$, 
$\theta\sim 1/\gamma$ is obtained for $\gamma <10^2$.
On the other hand, the angle is constant as $\theta\simeq 10^{-2}$ for $\gamma>10^2$.
The electrons moving along the $z$-axis 
are slightly deflected and tend to have angles $\gtrsim 10^{-2}$ from the $z$-axis.
Since we want to see the radiation from the strongly accelerated particles, we set the observer in 
the direction $\vec{n} = (n_x,n_y,n_z) = (10^{-2},0, \sqrt{1-10^{-4}})$.

\section{Discussion}
\label{discussion}
The superluminal strong waves (SLSWs) should exist around the termination shock of the pulsar wind nebulae.
They may play an important role in the particle acceleration.
Particularly, the particle acceleration by the SLSWs may work as a pre-acceleration mechanism for 
the Diffusive Shock Acceleration (DSA).
We note that for the electrons undergoing DSA, the contribution on the "energy change" is small 
since it is much slower than the Bohm limit of DSA.
To make the injection rate higher, the energy distribution should be broader.
The amplitude ratio between the primary and secondary waves is a key point for it.
If the upstream EM field mainly consists of primary SLSW, 
the resultant electron energy distribution tends to show a power law shape,
and the 4-velocity distribution is anisotropic.
On the other hand, if the SLSW distribution is isotropic and no primary wave exists, 
the obtained energy distribution is narrower than the former case and isotropic.
The SLSW acceleration in upstream may determine the injection rate of the DSA.
The radiation from the electrons accelerated by the SLSW can be understood as 
the synchro-Compton radiation in the secondary waves.
In this section, we discuss the applicability of this acceleration to the PWNe and the observational prospect.

\subsection{Length scale of the acceleration region}
The length scale the SLSWs can exist is also unsolved, but is an important problem.
Here we estimate the length scale needed for the acceleration.
We consider the observer frame ($K$ frame) and the upstream rest frame ($K'$ frame).
First, we consider the case for which these electrons are diffusively accelerated.
We assume the velocity direction is nearly parallel to the boost direction of the $K'$ frame in the $K$ frame.
The length scale in $K'$ frame for a typical deflection is $\sim c/\omega_{\rm min}'$,
where $\omega_{\rm min}'$ is the typical frequency of the waves.
This scale is written in the $K$ frame as $L \sim \Gamma (v' + V)/\omega_{\rm min}'$.
In the $K$ frame, we have
\begin{equation}
L \sim 4\Gamma^2 c/\omega_{\rm min},
\end{equation}
where $\omega_{\rm min} = \omega_0$ is the frequency in the $K$-frame which can be regarded as the 
inverse of the spin period of the pulsar.
It is estimated by using the Crab parameters as 
\begin{equation}
L \sim 6\times 10^{12} \left(\frac{\Gamma}{10^2}\right)^2\rm{cm}.
\end{equation}
Next we consider the electrons in the phase locking state.
The length scale in the $K'$ frame is $\sim a^2\gamma_0^2c/\omega_0'$,
where $\omega_0'$ is the frequency of the primary wave.
In the $K$ frame, it is 
\begin{equation}
L \sim 6\times 10^{16}\left(\frac{\Gamma}{10^2}\right)^2\left(\frac{\gamma_0'}{10}\right)^2\rm{cm}.
\end{equation}
This is shorter than the radius of the termination shock of Crab Nebula only by a factor of $5$.
From this estimation, it can be realized that the maximum Lorentz factor reached in our simulation is an upper limit,
since it may not oscillate even during in a period.
On the other hand, the planer wave approximation becomes inadequate if the length scale is
comparable to the termination shock radius.
Therefore, the acceleration by secondary waves without phase locking can play a role in the upstream, 
but we should be careful about the phase locking acceleration.
In other words, the maximum energy of the strong wave acceleration can be determined by the scale length 
the SLSWs exist.

\subsection{Observational prospect}
Lastly we discuss the possibility of observing the signature of the pre-acceleration by the primary wave.
The typical radiation in this situation is the synchro-Compton radiation from selectively accelerated electrons.
The maximum frequency of the synchro-Compton radiation in the $K'$ frame is written as
\begin{equation}
\omega_{\rm max} \sim \gamma_{\rm max}'^2\frac{e\varsigma}{mc}.
\end{equation}
The maximum Lorentz factor may be determined by the scale limit, as we have discussed above.
However, here we suppose the acceleration region is sufficiently large to reach
the maximum energy in one cycle of the phase locking electron $\gamma'_{\rm max}\sim a^2\gamma'_0$.
We note that the Photon Formation Length of the typical synchro-Compton photon (cf. Teraki \& Takahara 2014)
is much shorter than the length scale of the whole orbit.
Therefore, we can estimate the typical radiation frequency without the information of the whole orbit.
Here we suppose that the energy density of the SLSWs and electrons are highest near the region of the termination shock.
The typical frequency of this radiation in the observer frame is estimated as 
\begin{equation}
\omega \sim \Gamma \gamma_{\rm max}'^2\frac{eB/\Gamma}{mc} \\
\simeq2\times 10^{10}\left(\frac{\gamma'_{\rm max}}{10^3}\right)^2\left(\frac{B}{10^{-3}\rm{G}}\right)\rm{s}^{-1}.
\end{equation}
Thus, it can be observed as a radio knot.
The observable area is restricted as $\sim (r_{\rm TS}/\Gamma)^2$ by the beaming effect of the upstream bulk motion.
Interestingly, small radio knots with scale $\lesssim 10^{15}\rm{cm}$ are observed near the termination shock by VLBI imaging
(Lobanov, Horns, \& Muxlow 2011).
If we adjust the observable area to this observation, the bulk Lorentz factor is constrained to 
$\Gamma\gtrsim 10^2$.
This value is consistent with our scenario and acceptable for the PWNe models.
If these radio photons are emitted as in our model, unfortunately, no high energy counterparts 
emitted by the inverse Compton scattering will be observed.
The luminosity should be much smaller than the gamma-ray from the whole nebula,
and current gamma-ray observations cannot resolve spatial structure of the PWNe.

\section{Summary}
\label{summary}
We have investigated the electron acceleration in the superluminal strong waves and the radiation from them.
We considered two classes of waves.
One is the primary wave, and the other is the isotropically distributed secondary waves.
We took the amplitude ratio of them as a parameter.
When the primary wave is dominant, the electrons moving nearly along the wavevector direction are
selectively accelerated, and form a power law distribution in the high energy region.
On the other hand, when the secondary waves are dominant,
the energy distribution is narrow and shows an exponential cutoff, and does not show any power law tails beyond the peak.
We can expect both cases in the upstream of the termination shock of the PWNe.
If the former case is realized, this acceleration mechanism may play a significant role for the injection to the shock acceleration.
The radiation features can be described by synchro-Compton theory
when the amplitude of the secondary waves are similar to that of the primary wave.
The radiation from the phase locked electrons can show different spectrum
when the primary wave is extremely dominant and the strength parameter of the secondary waves is much smaller than unity.
However, such situation may not be realized around the termination shock of the PWNe, since
it is a very coherent situation.
Radio knots which are illuminated by synchro-Compton mechanism without counterparts in higher frequency range 
can be observed in the immediate upstream of the termination shocks if the SLSWs can propagate much longer than its wavelength.

\vspace{0.2in} 
We are grateful to the referee for his/her constructive and helpful comments.
We also thank Jin Matsumoto, Maria Dainotti, Maxim Barkov, Annop Wongwathanarat, Akira Mizuta, Tomoya Takiwaki,
and Nodoka Yamanaka for fruitful discussions. 
Y. T. thanks Fumio Takahara, Takanobu Amano, Shimone Giacche, and John Kirk for stimulating suggestions.
H. I. acknowledges support by Grant-in Aid for Young Scientists(B:26800159) from the Ministry of Education,
Culture Sports, Science and Technology, Japan.
This work is supported by the RIKEN through Special Postdoctoral Researcher Program.

\clearpage

\clearpage

\clearpage

\clearpage

\clearpage

\clearpage
\clearpage

\end{document}